\documentclass[a4article]{article}
\usepackage{graphicx}

\begin{document}
\title{The influence of electron density gradient direction on
gradient-drift instabilities in the E-layer of the ionosphere}

\author{O.I.Berngardt, A.P.Potekhin}
\maketitle

\begin{abstract}
We show that the dispersion relation for gradient-drift
and Farley-Buneman instabilities within the approximation of the two-fluid
MHD should contain the terms which are traditionally supposed to be
small. These terms are caused by taking into account divergence of
particles velocity and electron density gradient along the magnetic
field direction. It is shown that at heights below 115km the solution
of the dispersion relation transforms into standard one, except the
situations, when the electron density gradient is parallel to magnetic
field or wave-vector. In these cases the traditionally neglected summands
to the growth rate of the irregularities becomes significant.

The additional terms depend on relative directions of electron density
gradient, magnetic field and mean velocities. This leads to the different
instability growth conditions at equatorial and high-latitude regions
of the ionosphere. The obtained results do not contradict with the
experimental data.

\end{abstract}

\section{Introduction}

In the recent review \cite{Farley2009} the equatorial instabilities
at E-layer heights was analyzed, and the following problem was formulated:
\char`\"{}How different are the auroral and equatorial zones? Is the
physics essentially the same, except for the fact that the auroral
zone electric fields are often much stronger? Or is it important that
in the auroral zone there are gradients parallel to the magnetic field
in electron density, electron and ion temperatures and collision frequencies?\char`\"{}.
In this paper we try to answer to this question.

As stated in \cite{Farley2009}, the analysis of gradient-drift instabilities
in the ionospheric plasma at E-layer heights in linear approximation
is based on the theory suggested in \cite{Fejer_et_al_1975}. The
result of the work \cite{Fejer_et_al_1975} is the correct derival
of the dispersion relation for the plasma instabilities in the specific
case when electron density gradient is oriented vertically, magnetic
field - horizontally, electrons are magnetized and ions are unmagnetized.
In the work \cite{Fejer_et_al_1975} authors also retrieved the well known
solution of the dispersion relation. The attempt of solving the wider
problem, when electron density gradients and magnetic field have arbitrary
directions, was made in \cite{Fejer_et_al_1984} and is thought to
be valid both for equatorial and high-latitude ionosphere. Unfortunately
in the both papers the solution of the dispersion relation was obtained
under a number of additional assumptions. Due to this fact these theories
sometimes do not absolutely correctly describe the conditions of the
instabilities generation, especially when the instabilities wave-vector
or magnetic field are almost parallel to electron density gradient,
as mentioned in \cite{Farley2009}. The possible cause of these problems
is preliminary simplification of the dispersion relation and neglection
of important terms.

The goal of the paper is the analysis of the dispersion relation solution,
generalizing the formulas obtained in \cite{Fejer_et_al_1975,Fejer_et_al_1984}
by taking into account some terms traditionally neglected. We base
on the same two-fluid magnetohydrodynamics (MHD) approximation as
used in \cite{Fejer_et_al_1975} and assume that this approximation
is valid for the heights under investigation. We use approach suggested 
in \cite{Gershman1974,Gelberg1986} for studying the polar ionosphere. 
Then we investigate all the terms in the obtained solution to take into 
account important ones.

\section{Solution of dispersion relation}

\subsection{Traditonal approach}

The instability growth condition in presence of arbitrary direction
of electron density gradient and electric and magnetic fields in linear
approximation is retrieved, for example, in \cite{Fejer_et_al_1984}.
The dispersion relation obtained there has a form:


\begin{eqnarray}
\left(\omega+i2\alpha N_{0}-{\bf V_{d}}{\cdot}{\bf k}\right)\left(\nu_{in}-i\left(\omega-\frac{{\bf k_{\bot}}{\cdot}\left({\bf \nabla}N\times{\bf b}\right)\Omega_{i}}{k_{\bot}^{2}N}\right)\right) & +\nonumber \\
\left(\left(\omega+i2\alpha N_{0}\right)\left(\Omega_{i}^{2}+\left(\nu_{in}-i\omega\right)^{2}\right)+ik^{2}C_{s}\left(\nu_{in}-i\omega+i\frac{{\bf k_{\bot}}{\cdot}\left({\bf \nabla}N\times{\bf b}\right)\Omega_{i}}{k_{\bot}^{2}N}\right)\right)\cdot \label{eq:1} \\
\left(\frac{\Psi}{\nu_{in}}-i\frac{{\bf k_{\bot}}{\cdot}\left({\bf \nabla}N\times{\bf b}\right)\Omega_{i}}{k_{\bot}^{2}N}\right) & =0 \nonumber\end{eqnarray}

 where ${\bf V}_{d}={\bf V}_{e0}-{\bf V}_{i0}$ - relative electron/ion
drift velocity; $C_{s}=\sqrt{\frac{T_{e}+T_{i}}{m_{i}}}$ - ionacoustic
speed, $\alpha$ - recombination coefficient;

\begin{equation}
\Psi=\frac{\nu_{en}\nu_{in}}{\Omega_{e}\Omega_{i}}\left(1+\frac{\Omega_{e}^{2}k_{||}^{2}}{\nu_{en}^{2}k^{2}}\right)\label{eq:2}\end{equation}

 - a coefficient defining the aspect sensitivity of the instabilities,
${\bf b}=\frac{{\bf B}}{B}$.

To emphasize the problem of the dispersion relation (\ref{eq:1}),
let us analyze a special case, when electron density gradient is parallel
to the wave-vector and perpendicular to the magnetic field. In this
case we have 

\begin{equation}
{\bf k_{\bot}}{\cdot}\left({\bf \nabla}N\times{\bf b}\right)=0\label{eq:3}\end{equation}

and in the standard dispersion relation (\ref{eq:1}) the presence
of gradients can be neglected, and gradient-drift generation mechanism
stop working.

The condition (\ref{eq:3}) is almost valid for non-disturbed equatorial
ionosphere (where electron density gradient and wave vector are almost
vertical), and for non-disturbed high-latitude ionosphere (where electron
density gradient and magnetic field are almost vertical), so this
problematic situation should arise frequently.

But sometimes this contradicts with the experimental data, for example
at Earth magnetic equator, with vertical gradient of electron density
and vertical wave-vector of sounding wave (and vertical wave-vector
of irregularities, correspondingly) a powerful scattering is observed,
as stated in \cite[pp.1515-1516]{Farley2009}. This allows us to suggest
the presence of the gradient-drift instabilities even in this special
case.

It is obvious that in the process of the dispersion relation (\ref{eq:1})
output a number of approximations was used, and necessary effect may
be lost or neglected due to its small value. 

As a basis for our analysis we will use the approximation of
two-fluid MHD, following to the traditional approach developed in
\cite{Fejer_et_al_1975,Fejer_et_al_1984}. But in opposite to
this approach, following to the works \cite{Gershman1974,Gelberg1986},
we will not simplify the dispersion relation before obtaining the
solution.

\subsection{Full solution}

Usually, when investigating ionospheric instabilities, for example,
in \cite{Farley2009} it is thought that gradient-drift instabilities
should be investigated when ions are unmagnetized, and electrons are
magnetized, and velocity divergence effect can be neglected \cite{RogisterDAngelo1970}.
But it is well known that in fact the dispersion relation in two-fluid
MHD approximation has more general form, similar to the one obtained
in \cite{Fejer_et_al_1975}, but with taking into account the case
of arbitrary magnetized particles of both types (i.e., without preliminary
simplifications of the dispersion relation) and average velocity divergence
(see, for example, high-latitude case investigated in \cite{Gershman1974,Gelberg1986}).

It can be shown (see, for example, approach described in \cite[sect.5]{Gelberg1986})
that in this case the solution of the dispersion equation will be
more complex:


\begin{eqnarray}
\omega_{r}+i\Gamma & = & \frac{{\bf V}_{e0}+\Psi{\bf V}_{i0}}{1+\Psi}{\cdot}{\bf k}+i\frac{\Psi}{1+\Psi}\frac{1}{\nu_{in}^{t,\mu}}\left(\left(\frac{{\bf V_{0e}}+\Psi{\bf V_{0i}}}{1+\Psi}{\cdot}{\bf k}\right)^{2}-C_{s}^{2}k^{2}\right)+\nonumber \\
 &  & -i\left(2\alpha N_{0}\right)+i\frac{{\bf V}_{d}{\cdot}{\bf k}}{(1+\Psi)^{2}}\left(\Psi\frac{\Omega_{i}}{\nu_{in}^{t,\mu}}+\frac{\nu_{in}^{t,\mu}}{\Omega_{i}}\right)\left\{ \left(\frac{{\bf \bigtriangledown}N_{0}}{N_{0}k^{2}}\right){\cdot}\left({\bf b}\times{\bf k}\right)\right\} +\nonumber \\
 &  & -i\left(\frac{Q}{N_{\alpha0}}-\alpha N_{\alpha0}\right)+\nonumber \\
 &  & +i\frac{{\bf V}_{e0}+{\bf V}_{i0}\Psi}{1+\Psi}{\cdot}\frac{{\bf \bigtriangledown}N_{0}}{N_{0}}+\nonumber \\
 &  & -i\frac{\Psi{\bf V}_{d}{\cdot}{\bf k}}{(1+\Psi)^{2}}\frac{\Omega_{i}^{2}}{\left(\nu_{in}^{t,\mu}\right)^{2}k^{2}}\left(\frac{{\bf \bigtriangledown}N_{0}}{N_{0}}{\cdot}{\bf b}\right)\left({\bf b}{\cdot}{\bf k}\right)+\label{eq:4}\\
 &  & -i\frac{{\bf V}_{d}{\cdot}{\bf k}}{(1+\Psi)^{2}}\frac{m_{i}\nu_{e,n}^{t,\mu}}{m_{e}\nu_{i,n}^{t,\mu}k^{2}}\left(\left(\frac{{\bf \bigtriangledown}N_{0}}{N_{0}}{\cdot}{\bf k}\right)\frac{\left({\bf b}{\cdot}{\bf k}\right)^{2}}{k^{2}}-\left(\frac{{\bf \bigtriangledown}N_{0}}{N_{0}}{\cdot}{\bf b}\right)\left({\bf b}{\cdot}{\bf k}\right)\right)+\nonumber \\
 &  & +\frac{\frac{T_{e0}+T_{i0}}{m_{i}}}{\Omega_{i}(1+\Psi)}\left(\frac{\Omega_{i}^{2}}{\left(\nu_{in}^{t,\mu}\right)^{2}}\Psi-1\right)\left\{ \left(\frac{{\bf \bigtriangledown}N_{0}}{N_{0}}\right){\cdot}\left({\bf b}\times{\bf k}\right)\right\} \nonumber \end{eqnarray}

 where ${\bf V}_{d}={\bf V}_{e0}-{\bf V}_{i0}$ - drift velocity of
the electrons relative to ions. The solution was obtained from system
of two-fluid MHD equations for both types of particles (ions and electrons),
according to the consistency condition for the system, in long wavelength
approximation allowing to consider plasma density variations as quasi-neutral
ones ($\delta N_{e}=\delta N_{i}$).

It is important to note that average velocities ${\bf V}_{\alpha0}$
in traditional approximations of equal ionization and recombination
$Q-\alpha N_{\alpha0}^{2}=0$ and smooth velocity profile $\left({\bf V_{\alpha0}}{\bf \bigtriangledown}\right){\bf V_{\alpha0}}=0$
(see, for example, \cite{Golant1980}) can be defined as


\begin{equation}
{\bf V}_{\alpha0}=-\frac{\widehat{{\bf \sigma}}_{\alpha}}{Z_{\alpha}eN_{\alpha0}}({\bf E}_{0}-\frac{m_{\alpha}\nu_{\alpha n}^{t,\mu}}{Z_{\alpha}e}{\bf U}_{n})-\widehat{{\bf D}}_{\alpha}\frac{{\bf \bigtriangledown}N_{\alpha}}{N_{\alpha}}-\widehat{{\bf D}}_{T\alpha}\frac{{\bf \bigtriangledown}T_{\alpha}}{T_{\alpha}}\label{eq:5}\end{equation}

 where $\widehat{{\bf \sigma}}_{\alpha},\widehat{{\bf D}}_{\alpha},\widehat{{\bf D}}_{T\alpha}$
are the the operators of conductivity, diffusion and thermodiffusion
correspondingly, discussed, for example, in \cite{Golant1980}, and
$\nu_{\alpha n}^{t,\mu}=\frac{m_{n}}{m_{n}+m_{\alpha}}\nu_{\alpha n}^{t}$
- is a normed frequency of elastic collisions with neutrals.

In case, when $Q-\alpha N_{\alpha0}^{2}\neq0$ and/or $\left({\bf V_{\alpha0}}{\bf \bigtriangledown}\right){\bf V_{\alpha0}}\neq0$,
the velocities are defined by solving the following well known zero-order
equations numerically or analytically:


\begin{equation}
\left\{ \begin{array}{l}
{\bf V}_{\alpha0}{\cdot}{\bf \bigtriangledown}N_{\alpha0}+N_{\alpha0}{\bf \bigtriangledown}{\cdot}{\bf V}_{\alpha0}=Q-\alpha N_{\alpha0}^{2}\\
Z_{\alpha}eN_{\alpha0}{\bf E}_{0}+Z_{\alpha}eN_{\alpha0}{\bf V}_{\alpha0}\times{\bf B}_{0}+\\
\,+m_{\alpha}N_{\alpha0}({\bf V}_{\alpha0}{\cdot}{\bf \bigtriangledown}){\bf V}_{\alpha0}+{\bf \bigtriangledown}(T_{\alpha}N_{\alpha0})+N_{\alpha0}m_{\alpha}({\bf V}_{\alpha0}-{\bf U}_{n})\nu_{\alpha n}^{t,\mu}=0\\
0=Z_{i}N_{i0}+N_{e0}\end{array}\right.\label{eq:6}\end{equation}

It is important to note, that when investigating F-B and G-D instabilities
in two-fluid MHD approximation usually the following condition is
expected to be valid:


\begin{equation}
|Q-\alpha N_{\alpha0}^{2}|<<\alpha N_{\alpha0}^{2},Q\label{eq:7}\end{equation}

If that not the case, we should take into account densities of the
different kinds of particles (electrons and ions) in the ionization/recombination
term, which usually leads to the ionization waves, discussed, for
example, in \cite{Akhiezer_atal1974}.

\section{Discussion}

\subsection{Full solution}

By taking into account only significant terms at the E-layer altitudes, 
we obtain from equation (\ref{eq:4}) the following simplified solution 
for FB and GD instability:


\begin{eqnarray}
\omega_{r}+i\Gamma & = & \frac{{\bf V_{0e}}+\Psi{\bf V_{0i}}}{1+\Psi}{\cdot}{\bf k}+i\frac{\Psi}{1+\Psi}\frac{1}{\nu_{in}^{t,\mu}}\left(\left(\frac{{\bf V_{0e}}+\Psi{\bf V_{0i}}}{1+\Psi}{\cdot}{\bf k}\right)^{2}-C_{s}^{2}k^{2}\right)+ \nonumber\\
 &  & -i\left(2\alpha N_{0}\right)+i\frac{{\bf V}_{d}{\cdot}{\bf k}}{(1+\Psi)^{2}}\left(\frac{\nu_{in}^{t,\mu}}{\Omega_{i}}\right)\left\{ \left(\frac{{\bf \bigtriangledown}N_{0}}{N_{0}k^{2}}\right){\cdot}\left({\bf b}\times{\bf k}\right)\right\} +\nonumber \\
 &  & +i\frac{{\bf V}_{e0}+{\bf V}_{i0}\Psi}{1+\Psi}{\cdot}\frac{{\bf \bigtriangledown}N_{0}}{N_{0}}+\nonumber \\
 &  & +i\frac{{\bf V}_{d}{\cdot}{\bf k}}{(1+\Psi)^{2}}\frac{m_{i}\nu_{in}^{t,\mu}}{m_{e}\nu_{en}^{t,\mu}k^{2}}\left(\left(\frac{{\bf \bigtriangledown}N_{0}}{N_{0}}{\cdot}{\bf b}\right)\left({\bf b}{\cdot}{\bf k}\right)\right)\label{eq:8} \end{eqnarray}

The first five summands are well known:

first summand defines Farley-Buneman instability frequency in form,
obtained, for example, in \cite{Fejer_et_al_1975};

second one defines the growth rate of Farley-Buneman instability derived
in \cite{Farley1963,Buneman1963};

third one defines the affect of recombination processes to the growth
rate, obtained, for example, in \cite{Fejer_et_al_1975};

forth one defines the effect of electron density gradients
due to their arbitrary orientation to the magnetic field and wave-vector,
according to the paper \cite{Fejer_et_al_1984}. In a simple case
of gradients perpendicular to the magnetic field it transforms to
the widely used (as stated in \cite{Farley2009}) summand obtained
in \cite{Fejer_et_al_1975}.

fifth one defines velocity convergence effect. Usually the term is neglected. 
It is well known that the term actually can be neglected in case of gradient-free 
case \cite{RogisterDAngelo1970},
but in general case it should be taken into account, see, for example,
\cite{Gershman1974},\cite[eq.5.15]{Gelberg1986}.

In the equatorial ionosphere the last term can be neglected due
to the magnetic field and electron density gradients are almost orthogonal
and the term is small. But in general case, valid for both polar and
equatorial ionosphere, one should take it into account.

It is important to note that the first four summands in the solution
are well known, and analyzed, for example, in \cite{Fejer_et_al_1984},
but the last two summands affects only on growth rate and becomes
significant in the regions where electron density gradient is parallel
to the magnetic field or wave-vector, and traditional gradient term,
obtained in \cite{Fejer_et_al_1975,Fejer_et_al_1984} (forth summand
in the expression (\ref{eq:8}) ) becomes zero.

It is also important to note that the second summand, which defines
F-B mechanism of instability growth, decrease exponentially with altitude
due to decreasing the ratio $\frac{\Psi}{1+\Psi}\frac{1}{\nu_{in}^{t,\mu}}$
with increasing the height. From the other side, the fifth term does
not decrease so quickly so it depends mostly from the relative
gradient of electron density $\frac{{\bf \bigtriangledown}N_{0}}{N_{0}}$.
This leads to the importance of the fifth term at higher altitudes.

As one can see, the forth term, which defines traditional G-D decrement,
also decrease exponentially with altitude due to decreasing the ratio
$\left(\frac{\nu_{in}^{t,\mu}}{\Omega_{i}}\right)$. At the same time,
the sixth term does not have such a fast changes with altitude, due
to $\frac{m_{i}\nu_{in}^{t,\mu}}{m_{e}\nu_{en}^{t,\mu}k^{2}}$ relation
remains almost constant with altitude. So the sixth term becomes also
important at higher altitudes (above 100km).

Let us qualitatively analyze the obtained solution (\ref{eq:8}) by
supposing that electron density gradient is almost vertical. This
situation corresponds to the non-disturbed E-layer, produced mostly
by ionization/recombination processes. By doing this we neglect the
presence of turbulence and wave-like processes, which should be taking
into account during more detailed analysis.

\subsection{Equatorial case}

At the equator we can neglect the sixth term in (\ref{eq:8}), and
the solution becomes a bit simpler:


\begin{eqnarray}
\omega_{r}+i\Gamma & = & \frac{{\bf V_{0e}}+\Psi{\bf V_{0i}}}{1+\Psi}{\cdot}{\bf k}+i\frac{\Psi}{1+\Psi}\frac{1}{\nu_{in}^{t,\mu}}\left(\left(\frac{{\bf V_{0e}}+\Psi{\bf V_{0i}}}{1+\Psi}{\cdot}{\bf k}\right)^{2}-C_{s}^{2}k^{2}\right)+ \nonumber\\
 &  & -i\left(2\alpha N_{0}\right)+i\frac{{\bf V}_{d}{\cdot}{\bf k}}{(1+\Psi)^{2}}\left(\frac{\nu_{in}^{t,\mu}}{\Omega_{i}}\right)\left\{ \left(\frac{{\bf \bigtriangledown}N_{0}}{N_{0}k^{2}}\right){\cdot}\left({\bf b}\times{\bf k}\right)\right\} +\nonumber \\
 &  & +i\frac{{\bf V}_{e0}+{\bf V}_{i0}\Psi}{1+\Psi}{\cdot}\frac{{\bf \bigtriangledown}N_{0}}{N_{0}} \label{eq:9} \end{eqnarray}

Let us analyze the situation, when magnetic field direction $b$ is
horizontal, wavevector ${\bf k}$ is nearly vertical, vertical average
velocity ${\bf V_{0e\bot}}={\bf V_{0i\bot}}=20m/s$, horizontal electron/ion
drift velocity at 100-110 km $V_{d}=V_{0e}-V_{0i}=200m/s$ (the geometry
and physics of this is discussed, for example, in \cite{Farley2009}),
vertical gradient $\frac{{\bf \bigtriangledown}N_{0}}{N_{0}}$ is
supposed to be $10^{-4}m^{-1}$, $k=0.1m^{-1}$and almost vertical.
It is important to note, that in equatorial case the electron/ion
drift velocity ${\bf V}_{d}$ is strictly horizontal (see, for example,
\cite{Farley2009}). From the other side, the average velocity has
a vertical component, and the last term in (\ref{eq:9}) is always
not equal to zero.

At Fig.1 we show a height dependence of the second, forth and
fifth terms in (\ref{eq:9}) on the altitude for the described conditions.
Geometry is shown at Fig.1(H). Figures 1B-D corresponds to different
zenith angle in east-west plane (flow angle $\beta$, correspondingly
1.0,10.0,30.0 degrees, for zero aspect angle), figures E-G corresponds
to the different zenith angle in north-south plane (aspect angle $\alpha$,
correspondingly 0.1,1.0,3.0 degrees, for zero flow angle).

Figure 1A corresponds to both angles equal to zero. As one can see,
at all heights the velocity divergence term $\frac{{\bf V}_{e0}+{\bf V}_{i0}\Psi}{1+\Psi}{\cdot}\frac{{\bf \bigtriangledown}N_{0}}{N_{0}}$
(dot-dashed line) in this case is much larger than standard F-B increment
term $\frac{\Psi}{1+\Psi}\frac{1}{\nu_{in}^{t,\mu}}\left(\frac{{\bf V_{0e}}+\Psi{\bf V_{0i}}}{1+\Psi}{\cdot}{\bf k}\right)^{2}$
(dotted line) and due to this must be always taken into account. At
heights above 115-120 km it becomes larger than decrement term $\frac{\Psi}{1+\Psi}\frac{1}{\nu_{in}^{t,\mu}}C_{s}^{2}k^{2}$
(solid line) and this fact can cause an additional growth of the instabilities.
This term can be important, for example, for analysis of the echo
at heights above 115km (see, for example, \cite[ Fig.3]{Farley2009}).

As one can also see, at flow angles higher than 10 degrees (Figure
1C-D) standard GD term $\frac{{\bf V}_{d}{\cdot}{\bf k}}{(1+\Psi)^{2}}\left(\frac{\nu_{in}^{t,\mu}}{\Omega_{i}}\right)\left\{ \left(\frac{{\bf \bigtriangledown}N_{0}}{N_{0}k^{2}}\right){\cdot}\left({\bf b}\times{\bf k}\right)\right\} $
(dashed line) becomes important for generation of the instabilities,
and it is well known fact (see, for example, \cite{Farley2009}).
Also, as one can see, the effect of the new term will be stronger
at larger wavelengths and weaker at smaller wavelengths.

It can be easily shown, that for lower part of ionospheric E-region,
when electron density gradient and velocities are co-directed (this
corresponds to the day conditions of the equatorial ionosphere), the
last term increases the growth rate. In this case the possibility
of observation of the irregularities becomes higher. When the electron
density gradient and velocities are anti-directed (this corresponds
to the night conditions of the lower part of E-layer of the equatorial
ionosphere), the last term decreases the growth rate. In this case
the possibility of observation of the irregularities becomes lower.

This effect - increasing of irregularities level at day and decreasing
at night is observed at the 100-110 km. altitudes more-less regularly
(see, for example, in \cite{Farley1985,Abdu_etal2002,Denardini_etal2005}).
The similar hourly dependence of the irregularities is observed even
at higher altitudes at equator (see, for example, in \cite{ChauKudeki2006}).
It is important to note that daytime and nighttime power dynamics
at 100-110km is explained for non-perpendicular wavevector and mean
electron velocity by equatorial electrojet dynamics (see, for example,
in \cite{Farley2009}), similar dynamics of scattered power at higher
altitudes (above 130km) is still unexplained (see, for example, in
\cite{ChauKudeki2006}). So the new term have similar dynamics and
does not contradict with these known experimental results.

\subsection{Quasi-polar case}

Let us analyze the following case: strong geomagnetic disturbance
produces almost horizontal ${\bf V_{0e}}={\bf V_{d}}=400m/s,{\bf V_{0i}}=0$,
electron density gradient and magnetic field are nearly parallel (within
10 degrees) and almost vertical, wavevector nearly perpendicular to
the magnetic field, $\frac{{\bf \bigtriangledown}N_{0}}{N_{0}}=10^{-4}m^{-1}$and
vertical, wavevector $k=0.1m^{-1}$ and almost horizontal. In this
case we can see that the last summand in the solution (\ref{eq:8})
is significant and the solution (\ref{eq:8}) becomes:


\begin{eqnarray}
\omega_{r}+i\Gamma & = & \frac{{\bf V_{0e}}}{1+\Psi}{\cdot}{\bf k}+i\frac{\Psi}{1+\Psi}\frac{1}{\nu_{in}^{t,\mu}}\left(\left(\frac{{\bf V_{0e}}}{1+\Psi}{\cdot}{\bf k}\right)^{2}-C_{s}^{2}k^{2}\right)+ \nonumber\\
 &  & -i\left(2\alpha N_{0}\right)+i\frac{{\bf V}_{0e}{\cdot}{\bf k}}{(1+\Psi)^{2}}\left(\frac{\nu_{in}^{t,\mu}}{\Omega_{i}}\right)\left\{ \left(\frac{{\bf \bigtriangledown}N_{0}}{N_{0}k^{2}}\right){\cdot}\left({\bf b}\times{\bf k}\right)\right\} +\nonumber \\
 &  & +i\frac{{\bf V}_{0e}{\cdot}{\bf k}}{(1+\Psi)^{2}}\frac{m_{i}\nu_{in}^{t,\mu}}{m_{e}\nu_{en}^{t,\mu}k^{2}}\left(\left(\frac{{\bf \bigtriangledown}N_{0}}{N_{0}}{\cdot}{\bf b}\right)\left({\bf b}{\cdot}{\bf k}\right)\right) \label{eq:10} \end{eqnarray}

As one can see, the last summand in expression (\ref{eq:10}) can
produce instabilities at aspect angles larger than traditional theory
predicts. The \char`\"{}large aspect angles\char`\"{} effect is known
in high latitude ionosphere and has different explanations (see, for
example, in \cite{Hamza_StMaurice_1995} and references there). As
also one can see, the new term do not depend on wavenumber, and standard
F-B term increases with wavenumber growth. Due to this the presence
of the last term will be more significant at larger wavelengths than
at shorter ones.

At Fig.2 the numerical results are shown. Electron density gradient
is supposed to be vertical, magnetic field does not have east-west
component and is about 10 degrees off the zenith. Maximal drift velocity
is 400m/s and oriented in north-south direction, which corresponds
to average geomagnetic disturbance. Geometry is shown at Fig.2(H).
Fig.2B-D shows dependence in the angle in plane perpendicular to the
magnetic field (flow angle $\beta$, 1.0,3.0,10.0 degrees correspondingly,
aspect angle 1.0 degree), Fig.2E-G shows dependence in plane of magnetic
field line (aspect angle $\alpha$, 1.0,3.0,5.0 degrees correspondingly,
zero flow angle). Fig 2A corresponds to the zero flow and aspect angles.
As one can see, the new term (dot-dashed line) is not important at
zero aspect angles, but becomes important at aspect angles 1-3 degrees
(Fig.2E-F), and at these aspect angles can be responsible for generation
of instabilities, when standard F-B increment term $\frac{\Psi}{1+\Psi}\frac{1}{\nu_{in}^{t,\mu}}\left(\frac{{\bf V_{0e}}+\Psi{\bf V_{0i}}}{1+\Psi}{\cdot}{\bf k}\right)^{2}$
(dotted line) is smaller than decrement term $\frac{\Psi}{1+\Psi}\frac{1}{\nu_{in}^{t,\mu}}C_{s}^{2}k^{2}$
(solid line). Due to specific orientation of the electron density
gradient and magnetic field, the standard G-D term $\frac{{\bf V}_{d}{\cdot}{\bf k}}{(1+\Psi)^{2}}\left(\frac{\nu_{in}^{t,\mu}}{\Omega_{i}}\right)\left\{ \left(\frac{{\bf \bigtriangledown}N_{0}}{N_{0}k^{2}}\right){\cdot}\left({\bf b}\times{\bf k}\right)\right\} $
(dashed line), is almost unimportant at these small flow angles.

As one can see, the new term can cause generation of the instabilities
at larger aspect angles (Figs.2E-F).

\section{Conclusion}

Our analysis has shown that growth rate for gradient-drift instabilities
has significant terms that depends on relative direction of electron
density gradients, mean velocity and magnetic field directions (\ref{eq:8}). 
This leads to the different conditions for the growth of the F-B and G-D
instabilities in the equatorial and auroral zones of the ionosphere.
These terms are usually neglected \cite{Fejer_et_al_1975,Fejer_et_al_1984,Farley2009}, 
but for building a general theory they must be taken into account.

At heights below 115km the full solution (\ref{eq:8}) transforms into standard
one \cite{Fejer_et_al_1984}, except the situations, when the electron density gradient is
parallel to magnetic field or wave-vector. In these cases the new
terms in the growth rate of the instabilities becomes significant
and predicts the presence of gradient-drift instabilities even in
these cases, as shown by expression (\ref{eq:8}), in opposite to
traditional theory obtained in \cite{Fejer_et_al_1975,Fejer_et_al_1984}.
At low-latitude region these terms may cause additional growth of
the instabilities at higher altitudes (above 115-120km). At high-latitude
region these terms may produce changes of irregularities aspect sensitivity
(large aspect angles effect). Both effects do not contradict with
experimental data.

\section*{Acknowledgements}

The work was done under
financial support of RFBR grant \#07-05-01084a.

\newpage
\begin{figure}
\includegraphics[scale=0.5]{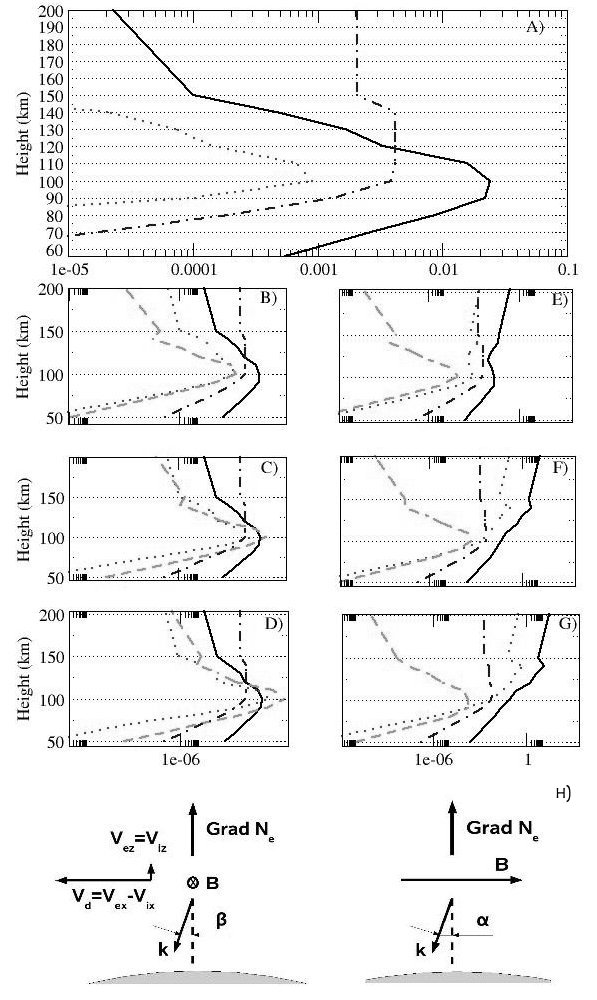}

\caption{The absolute value of each term in the solution (\ref{eq:9}) in
the equatorial ionosphere for different aspect ($\alpha$) and flow
($\beta$) angles. Figures (B-D) corresponds to different flow angles,
correspondingly 1.0,10.0,30.0 degrees, for zero aspect angle. Figures
(E-G) corresponds to the different aspect angles, correspondingly
0.1,1.0,3.0 degrees, for zero flow angle. Figure (A) corresponds to
both angles are equal to zero. Terms: $\frac{\Psi}{1+\Psi}\frac{1}{\nu_{in}^{t,\mu}}C_{s}^{2}k^{2}$
- solid line; $\frac{\Psi}{1+\Psi}\frac{1}{\nu_{in}^{t,\mu}}\left(\frac{{\bf V_{0e}}+\Psi{\bf V_{0i}}}{1+\Psi}{\cdot}{\bf k}\right)^{2}$
- dotted line; $\frac{{\bf V}_{d}{\cdot}{\bf k}}{(1+\Psi)^{2}}\left(\frac{\nu_{in}^{t,\mu}}{\Omega_{i}}\right)\left\{ \left(\frac{{\bf \bigtriangledown}N_{0}}{N_{0}k^{2}}\right){\cdot}\left({\bf b}\times{\bf k}\right)\right\} $
- dashed line; $\frac{{\bf V}_{e0}+{\bf V}_{i0}\Psi}{1+\Psi}{\cdot}\frac{{\bf \bigtriangledown}N_{0}}{N_{0}}$
- dot-dashed line. At Fig. (H) the geometry and explanation of $\alpha$ and $\beta$ is shown.}
\end{figure}

\newpage
\begin{figure}
\includegraphics[scale=0.5]{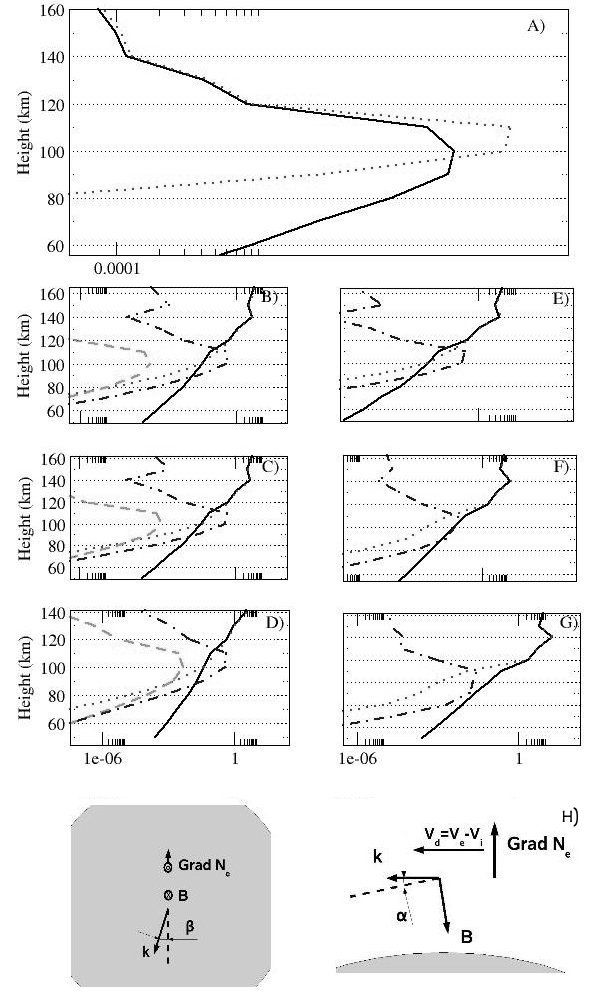}

\caption{The absolute value of each term in the solution (\ref{eq:10}) in
the high-latitude ionosphere for different aspect ($\alpha$) and
flow ($\beta$) angles. Figures (B-D) corresponds to different flow
angles , correspondingly 1.0,3.0,10.0 degrees, for 1.0 degree aspect
angle. Figures (E-G) corresponds to the different aspect angles, correspondingly
1.0,3.0,5.0 degrees, for zero flow angle. Figure (A) corresponds to
both 1.0 degree flow angle and zero aspect angle. Terms: $\frac{\Psi}{1+\Psi}\frac{1}{\nu_{in}^{t,\mu}}C_{s}^{2}k^{2}$
- solid line; $\frac{\Psi}{1+\Psi}\frac{1}{\nu_{in}^{t,\mu}}\left(\frac{{\bf V_{0e}}+\Psi{\bf V_{0i}}}{1+\Psi}{\cdot}{\bf k}\right)^{2}$
- dotted line; $\frac{{\bf V}_{d}{\cdot}{\bf k}}{(1+\Psi)^{2}}\left(\frac{\nu_{in}^{t,\mu}}{\Omega_{i}}\right)\left\{ \left(\frac{{\bf \bigtriangledown}N_{0}}{N_{0}k^{2}}\right){\cdot}\left({\bf b}\times{\bf k}\right)\right\} $
- dashed line; $\frac{{\bf V}_{0e}{\cdot}{\bf k}}{(1+\Psi)^{2}}\frac{m_{i}\nu_{in}^{t,\mu}}{m_{e}\nu_{en}^{t,\mu}k^{2}}\left(\left(\frac{{\bf \bigtriangledown}N_{0}}{N_{0}}{\cdot}{\bf b}\right)\left({\bf b}{\cdot}{\bf k}\right)\right)$
- dot-dashed line. At Fig. (H) the geometry and explanation of $\alpha$ and $\beta$ is shown.}
\end{figure}



\begin{thebibliography}{10}

\bibitem[Abdu et al.,2002]{Abdu_etal2002}M.A.Abdu, C.M.Denardini, J.H.A.Sobral, I.S.Batista,
P.Muralikrishna, E.R. dePaula, //{\it Journ. Atmosph. and Sol.-Terr.Phys.} {\it 64},
1425-1434, 2002

\bibitem[Akhiezer et al,1974]{Akhiezer_atal1974} A.I. Akhiezer, I.A. Akhiezer, R.V. Polovin,
A.G. Sitenko, and K.N. Stepanov. \textit{Plasma Electrodynamics,} Volumes 1-2.
Oxford: Pergamon Press, 1975

\bibitem[Buneman 1963]{Buneman1963}O.Buneman , Excitation of field
aligned sound waves by electron streams//
{\it Phys.Rev.Lett.} {\it 10}, 285-287, 1963

\bibitem[Chau and Kudeki,2006]{ChauKudeki2006}J.L.Chau , E.Kudeki
, Statistics of 150km echoes over Jicamarca based on low-power VHF
observations //{\it Ann.Geophys.} {\it 24}, 1305-1310, 2006

\bibitem[Denardini et al.,2005]{Denardini_etal2005}C.M.Denardini, M.A.Abdu, E.R. dePaula,
J.H.A.Sobral, C.M.Wrasse, Seasonal characterization of the equatorial electrojet height 
rise over Brasil as observed by RESCO 50MHz back-scatter radar //{\it Journ. Atmosph. and Sol.-Terr.Phys.}
 {\it 67}, 1665-1673, 2005


\bibitem[Farley,1963]{Farley1963}D.T.Farley, A plasma instability
resulting in field-aligned irregularities in the ionosphere, //{\it JGR }
{\it 68}, 6083-6097, 1963

\bibitem[Farley,1985]{Farley1985}D.T.Farley, Theory of equatorial electrojet plasma waves: 
new developments and current system //{\it JATP} {\it 47}, 729-744,
1985


\bibitem[Farely,2009]{Farley2009}D.T.Farley , The equatorial E-region
and its plasma instabilities: a tutorial //{\it Ann.Geophys.} {\it 27} 1509-1520,
2009

\bibitem[Fejer et al.,1975]{Fejer_et_al_1975}B.G.Fejer , D.T.Farley ,
B.B.Balsley  and R.F.Woodman , Vertical Structure of the VHF Backscattering
Region in Equatorial Electrojet and the Gradient Drift Instability,
//{\it JGR} {\it 80} 1313-1324 , 1975

\bibitem[Fejer et al.,1984]{Fejer_et_al_1984}B.G.Fejer, J.Providakes
, D.T.Farley , Theory of plasma waves in the auroral E region,
//{\it JGR} {\it 89} 7487-7494, 1984

\bibitem[Gelberg,1986]{Gelberg1986}M.G.Gelberg, \textit{Irregularities of high-latitude
ionosphere (In Russian)}, Novosibirsk, Nauka, 192pp., 1986

\bibitem[Golant et al.,1980]{Golant1980} V.E.Golant , P.Zhilinsky  and
I.E.Sakharov, \textit{Fundamentals of Plasma Physics,} 405pp., 1980

\bibitem[Gershman,1974]
{Gershman1974} B.N.Gershman , \textit{Dynamics of
the ionospheric plasma (In Russian)}, Moscow: Nauka, 256pp., 1974


\bibitem[Hamza and St.Maurice,1995]{Hamza_StMaurice_1995}A.M.Hamza and J.-P. St.Maurice,
Large aspect angles in auroral E-region echoes: a self-consistent turbulent fluid theory,
//{\it JGR} {\it 100} 5723-5732, 1995



\bibitem[Rogister and D'Angelo,1970]{RogisterDAngelo1970} Rogister,D'Angelo, Type II irregularities in the Equatorial electrojet
//{\it JGR} {\it 75}, 3879-3887, 1970











\end{thebibliography}
\end{document}